\documentclass[reqno]{amsart} 
\usepackage{amssymb}
\usepackage{amsmath}
\usepackage{amsfonts}

\newtheorem{theorem}{Theorem}

\theoremstyle{remark}

\newcommand{\nc}{\newcommand}
\newcommand{\Z}{\mathbb Z}
\nc{\rnc}{\renewcommand} \nc{\beq}{\begin{equation}}
\nc{\eeq}{\end{equation}} \nc{\beqa}{\begin{eqnarray}}
\nc{\eeqa}{\end{eqnarray}} \nc{\T}{\mathbb{T}}

\def\Tr{\operatorname{Tr}}
\def\rank{\operatorname{rank}}
\def\ind{\operatorname{ind}}
\def\Sym{\operatorname{Sym}}

\begin{document}

\begin{flushright} AEI-2011-015 \end{flushright}

\title[$3d$ superconformal indices]
{Superconformal indices of
three-dimensional theories related by mirror symmetry}

\author[C. Krattenthaler]{C. Krattenthaler$^\dagger$}
\address{Fakult\"at f\"ur Mathematik, Universit\"at Wien,
Nordbergstra$\beta$e 15, A-1090 Vienna, Austria}

\author[V.~P.~Spiridonov]{V.~P.~Spiridonov$^\ddagger$}
\address{Bogoliubov Laboratory of Theoretical Physics,
JINR, Dubna, Moscow Region 141980, Russia}

\author{G.~S.~Vartanov}
\address{Max-Planck-Institut f\"ur Gravitationsphysik, Albert-Einstein-Institut
14476 Golm, Germany; e-mail address: vartanov@aei.mpg.de}

\thanks{$^\ddagger$Research partially supported by the Austrian
Science Foundation FWF, grants Z130-N13 and S9607-N13,
the latter in the framework of the National Research Network
``Analytic Combinatorics and Probabilistic Number Theory."\newline\indent
$\ddagger$Research partially supported by
RFBR (grant no.~09-01-00271).}

\begin{abstract}
Recently, Kim, and Imamura and Yokoyama derived an exact formula for
superconformal indices in three-dimensional field theories. Using
their results, we prove analytically the equality of superconformal
indices in some $U(1)$-gauge group theories related by
mirror symmetry. The proofs are based on well-known identities
in the theory of $q$-special functions. We also suggest
a general index formula taking into account the $U(1)_J$
global symmetry present for abelian theories.
\end{abstract}

\maketitle


\section{Introduction}

The superconformal index (SCI) technique is a very useful tool for
testing supersymmetric dualities. Initially, this technique was
introduced for four-dimensional supersymmetric field theories
\cite{Kinney,Romelsberger1} in the context of $\mathcal{N}=1$ SYM
Seiberg dualities and AdS/CFT correspondence for $\mathcal{N}=4$ SYM
field theories. In this case, the indices are described by the
elliptic hypergeometric integrals \cite{S1} as observed first by
Dolan and Osborn \cite{Dolan:2008qi}. Various developments and
applications of this technique are described in
\cite{SV,SV1,V,Spiridonov:2010em,GPRR1}.

Later, in \cite{Bhattacharya:2008zy}, the superconformal index for
three-dimensional supersymmetric Chern--Simons theories  with large
rank of the gauge group $N$ was introduced, and the coincidence with
the gravitational background index in the context of AdS/CFT
correspondence \cite{Aharony:2008ug} has been established.
In \cite{Dolan:2008vc} superconformal characters of three-dimensional
supersymmetric theories have been constructed and, after taking
restrictions for parameters in them, SCIs for theories considered in
\cite{Bhattacharya:2008zy} can be obtained in a different way.
In \cite{Choi:2008za,Kim:2009wb,Kim:2010vwa,Imamura:2011uj,Cheon:2011th},
various three-dimensional SCIs were calculated in the large $N$ limit
for comparison with their gravity duals. The partition functions of
three-dimensional supersymmetric field theories are studied in
\cite{Kapustin:2010xq,Drukker:2010nc,Herzog:2010hf,
Jafferis:2010un,Hama:2010av,Martelli:2011qj,Cheon:2011vi,Jafferis:2011zi}.

The superconformal index for $\mathcal{N}=6$ Chern--Simons theory
with \textit{finite $N$} was derived by Kim in \cite{Kim:2009wb}.
The contribution
to SCIs of chiral fields with arbitrary $R$-charge was found
recently by Imamura and Yokoyama in \cite{Imamura:2011su}.
After combining everything, this gives an exact formula for $3d$
SCIs analogous to R\"omelsberger's result for $4d$
$\mathcal{N}=1$ SYM theories \cite{Romelsberger1}. SCIs of
some $3d$ $\mathcal{N}=2$ supersymmetric field theories
and their mirror partners \cite{Intriligator:1996ex} (see also
\cite{deBoer:1997ka,Aharony:1997bx} for a general discussion of such
theories) were computed in \cite{Imamura:2011su} and their coincidence
was confirmed up to the first several terms of the corresponding
series expansions in chemical potentials.
The main goal of the present work consists in the analytic proof
of exact coincidence of SCIs for these mirror symmetric
$3d$ theories.

Let us describe the SCI derived in \cite{Kim:2009wb,Imamura:2011su}.
For two particular supercharges $\mathcal{Q}$ and
$\mathcal{Q}^\dag=\mathcal{S}$, the superconformal partner of
$\mathcal{Q}$, the following algebraic relation holds:
 \beq
\label{anticom} \{ \mathcal{Q},\mathcal{Q}^\dag
\} \ = \ 2 \mathcal{H} = \Delta-R-J_3,
\eeq
where the space-time symmetry generators $\Delta, R,
J_3$ are the Hamiltonian, the $R$-charge, and the third component of
the angular momentum, respectively. One defines an extension
of the Witten index for the  theories compactified on the two-sphere
$S^2$ as
\beq
I \ = \ \Tr \left(
(-1)^{\mathcal{F}} e^{-\beta \mathcal{H}} x^{\Delta+J_3}
\prod_{i=1}^{\rank F} t_i^{F_i} \right),
\label{trace}\eeq
where $\mathcal{F}$ is the fermion number operator
and $x$ is the chemical potential
associated with the operator $\Delta+J_3$ commuting with the
chosen supercharges. Chemical potentials $t_i$ are
associated with commuting generators of the group $F$ describing
other global symmetries of the theory. Analogously
to the $4d$ case one computes the trace over the space of
states defined by the kernel of $\mathcal{Q}$ and $\mathcal{Q}^\dag$,
 since contributions of the states with non-zero eigenvalues of
$\mathcal{H}$ cancel each other. In other words, one works with
the gauge invariant BPS states protected by at least one pair of
supercharges which do not form long multiplets. As a result,
the index does not depend on $\beta$.

Similar to the $4d$ case, the $3d$ SCI is computed in two steps by
using the localization procedure in $\mathbf{S}^2 \times
\mathbf{S}^1$. One constructs first the so-called single-particle
state index and then computes the full SCI. The single-particle
state index is defined by the formula
\cite{Kim:2009wb,Imamura:2011su}
\begin{multline} \label{SCI_s}
\ind(e^{\textup{i}g_j},s,x,t)
= -\sum_{\alpha\in G}
e^{\textup{i}\alpha(g )}x^{2|\alpha(s )|}
\\
 + \sum_\Phi \sum_{\rho\in R_\Phi} \left[
e^{\textup{i} \rho(g )}t_i^{f_i}\frac{x^{2|\rho(s )|+\Delta_\Phi}}{1-x^2}
-e^{-\textup{i} \rho(g )}t_i^{-f_i}
\frac{x^{2|\rho(s )|+2-\Delta_\Phi}}{1-x^2} \right],
\end{multline}
where the first term describes contributions of the gauge fields,
and the rest comes from the matter fields with flavour charges $f_i$.
The chemical potentials $g =\{ g_1,\ldots,$ $g_{\rank G} \}$ are
associated with the generators of the maximal torus of the gauge group $G$.
Similarly we write $s =\{ s_1,\ldots,s_{\rank  G} \}$,
where $s_j$ are some half-integers associated with the magnetic monopole fluxes.
The sum $\sum_{\alpha\in G}$ is taken
over the roots of the Lie algebra of $G$.
Without the term $x^{2|\alpha(s )|}$ it would yield essentially the
character for the adjoint representation.
This contribution was computed in \cite{Kim:2009wb}.

The second term was also computed in case of $\mathcal{N}=6$
superconformal Chern--Simons theory in \cite{Kim:2009wb} for the
matter fields with the specific $R$-charges $\Delta_{\Phi}=1/2$. In
\cite{Imamura:2011su}, the contribution of chiral fields with
general $R$-charges is determined. Here $\Delta_{\Phi}$ is the Weyl
weight of a chiral multiplet $\Phi$ lying in the representation
$R_{\Phi}$ of the gauge group $G$. Similar to the $4d$ case, the
scalar component of the chiral superfield $\Phi$ has $R$-charge
equal to $\Delta_{\Phi}$, and the fermion component has the
$R$-charge $\Delta_{\Phi}-1$. The sum $\sum_{\rho\in R_\Phi}$ is the
sum over all terms with the weight $\rho(g)$ for a given chiral
field $\Phi$ lying in the $R_{\Phi}$ representation of the gauge
group $G$. The symbols $\alpha(g)$ and $\alpha(s)$ in the first term
are used for a separate presentation of the gauge field and the
monopole contributions coming from the adjoint representation of
$G$.

Having the single-particle state index, one derives a full SCI using
the plethystic exponent \cite{Imamura:2011su}
\begin{multline} \label{SCI_f}
I(x,t)=\sum_{s} \frac{1}{\Sym} \int
e^{-S_{\rm CS}^{(0)}} e^{\textup{i} b_0(g)} x^{\epsilon_0}
\prod_{i=1}^{\rank F} t_i^{q_{0i}} \\ \cdot
\exp\left[\sum_{n=1}^\infty \frac{1}{n} \ind(z_j^n,s,x^n,t_i^n)
\right] \prod_{j=1}^{\rank  G}\frac{dz_j}{2 \pi \textup{i} z_j},
\end{multline}
where $z_j=e^{\textup{i} g_j}$,
\beq \epsilon_0=\sum_\Phi
(1-\Delta_\Phi)\sum_{\rho\in R_\Phi}|\rho(s)| -\sum_{\alpha\in
G}|\alpha(s)| \eeq is the zero-point contribution to the energy,
 \beq q_{0i}=-\sum_\Phi\sum_{\rho\in
R_\Phi}|\rho(s)|f_i \eeq
is the zero-point contribution to the flavour charges,
and, finally,
\beq
b_0(g)=-\sum_\Phi\sum_{\rho\in R_\Phi}|\rho(s)|\rho(g),
\eeq
which can be regarded as the one-loop correction to the Chern--Simons term.

In the presence of the Chern--Simons term, there is a
contribution $e^{-S_{\rm CS}^{(0)}}$, where
\beq
S_{\rm CS}^{(0)} = 2 \textup{i} \Tr _{CS} (gs),
\eeq
with $\Tr _{CS}$ standing for the trace including
Chern--Simons levels \cite{Kim:2009wb,Imamura:2011su}. For
$U(N)_k$ theory with Chern--Simons level $k$, this
factor equals $2\textup{i} k \sum_{j=1}^{\rank  G} s_j g_j$
\cite{Kim:2009wb}.

In contrast to the four-dimensional case, where the index contains
integration over the gauge group only \cite{Kinney,Romelsberger1,Dolan:2008qi},
 here one has additionally the sum over the fluxes of $\rank  G$ independent
monopoles. Monopoles appear as solutions of the classical field equation
associated with the saddle points in the localization procedure.
Effectively, this leads to the shifted spin $j$ and the shifted value of
eigenvalue for $J_3$ due to the contribution from the background fluxes
$m_j$, and the variable
$$
s_j=\frac{1}{2}m_j,\quad m_j\in\mathbb{Z},
$$
is introduced for convenience. In the above formulas, the
quantities $\rho(s)$
represent the same Weyl weights in this background flux for a taken
chiral field $\Phi$. For example, the chiral superfields ---
gauge group singlets have $\rho(g )=\rho(s )=0$.

As pointed out in \cite{Kim:2009wb}, the integration over the gauge group
is a little bit tricky, because of the presence of monopoles. It was
shown in \cite{Kim:2009wb} that the contribution coming from the
vector multiplet for $s_i \neq s_j$ and for $s_i = s_j$ is different,
since in the first case the monopole spherical harmonics and in the
latter case the usual spherical harmonics are used. As suggested in
\cite{Imamura:2011su}, this fact is already included in the term for
the contribution of the vector multiplet. The term
$\frac{1}{\Sym}$ in (\ref{SCI_f}) appears because of the same reason,
it is connected with the fact that the initial gauge group $G$ is
'broken' by the monopoles into the product $G_1 \times \dots \times
G_k$, which gives $\Sym = \prod_{i=1}^k (\rank  G_i)!$
\cite{Kim:2009wb}, which can also be written in the form
\beq
\Sym = \prod_{i=1}^{\rank G} \left( \sum_{j=i}^{\rank G} \delta_{s_i,s_j} \right),
\eeq
where $\delta_{a,b}$ is the usual Kronecker delta-function.

In \cite{Imamura:2011su}, SCIs for mirror symmetric theories
were calculated with some restrictions --- the same chemical
potentials were used for both quark fields belonging to one flavour.
Moreover, corresponding formulas did not contain the chemical potential
associated with the abelian symmetry group $U(1)_J$ \cite{Aharony:1997bx}
resolving degeneracies. Shifting the scalar
component of vector multiplet by an arbitrary constant, one can take
into account this extra global symmetry. In the Appendix, we present
SCIs with the most general set of chemical potentials.
Surprisingly, we found that the SCI terms
related to the symmetry group $U(1)_J$ can not be obtained
directly from the results of \cite{Imamura:2011su},
suggesting that they may be incomplete.

In contrast to the four-dimensional case, where in most cases
the condition of the anomaly absence fixes the $R$-charge, in
the three-dimensional case the $R$-charge in (\ref{SCI_f}) is not fixed.
Moreover, it can be arbitrary, since adding to it
any combination of
the abelian global charges represents again
the $R$-charge. The latter fact is reflected by the appearance
of the free parameter $h$ associated with the $R$-charge in the SCIs.
Recently the $Z$-extremization procedure was suggested in
\cite{Jafferis:2010un} for obtaining the exact $R$-charges of matter
fields in the IR fixed points of $3d$ theories.

Formulas (\ref{SCI_s}) and (\ref{SCI_f})
resemble to some extent the procedure of calculating SCIs in $4d$
supersymmetric field theories \cite{Kinney,Dolan:2008qi}, but they
are much more involved. In particular, there
are the terms in addition to the plethystic exponent, which is a
new structural element. The building block of SCIs for $3d$
theories is given by the infinite $q$-product
$$
(z;q)_\infty = \prod_{j=0}^\infty (1-zq^j),\qquad |q|<1.
$$

Let us consider the chiral superfield $\Phi$ with arbitrary $R$-charge
$\Delta_{\Phi}$ in the fundamental representation of $U(N_c)$.
Then the single-particle state index is \beq
\ind_C(x,e^{\textup{i}g_j},s_j) \ = \ \sum_{j=1}^{N_c}
\frac{x^{\Delta_{\Phi}+2|s_j|} e^{\textup{i} g_j} - x^{2-\Delta_{\Phi}+2|s_j|} e^{-\textup{i} g_j}}{1-x^2},\eeq
and the full SCI is obtained as
\begin{align} \notag
\exp \Bigg( \sum_{n=1}^\infty \frac 1n
\ind_C&(x^n,e^{\textup{i}ng_j},s_j) \Bigg)  \\
\notag
&= \exp \left( \sum_{n=1}^\infty \frac 1n \sum_{j=1}^{N_c}
\frac{x^{n(\Delta_{\Phi}+2|s_j|)} e^{\textup{i} ng_j}
 - x^{n(2-\Delta_{\Phi}+2|s_j|)} e^{-\textup{i} ng_j}}{1-x^{2n}} \right)
\\
&=
\prod_{j=1}^{N_c} \frac{(x^{2-\Delta_{\Phi}+2|s_j|} e^{-\textup{i}
    g_j};x^2)_\infty}{(x^{\Delta_{\Phi}+2|s_j|} e^{\textup{i}
    g_j};x^2)_\infty}.
\end{align}

The contribution of gauge fields looks fundamentally different.
When the gauge group is $U(N_c)$, the contribution of the vector multiplet
is \cite{Kim:2009wb}
\beq
\ind_V (x,e^{\textup{i} g_j},s_j) = - \sum_{i,j=1,\, i \neq j}^{N_c}
x^{|s_i-s_j|} e^{\textup{i} (g_i-g_j)},
\eeq
and the contribution to the full SCI is given by
\beq
\exp \left( \sum_{n=1}^\infty \frac 1n
\ind_V(x^n,e^{\textup{i} ng_j},s_j) \right) = \prod_{i,j=1,\,  i \neq j}^{N_c}
(1 - x^{|s_i-s_j|} e^{\textup{i} (g_i-g_j)}).
\eeq

We need also some mathematical definitions from the theory of
$q$-special functions (see, e.g., \cite{GaRaAA}). The basic
hypergeometric series ${}_{r+1}\phi_r$ is defined by
\beq {}_{r+1}\phi_r \!\left[ \begin{array}{c}
         a_1,a_2,\ldots,a_{r+1} \\
         b_1,b_2,\ldots,b_r
       \end{array}
;q,z \right] = \sum_{n=0}^{\infty} \frac{(a_1,a_2, \ldots,
a_{r+1};q)_n}{(q,b_1,b_2,\ldots,b_r;q)_n} z^n,\eeq
where
$$
(a_1,a_2, \ldots, a_{r+1};q)_n:=(a_1;q)_n(a_2;q)_n \cdots
(a_{r+1};q)_n,
$$
with the $q$-shifted factorial $(z;q)_n$ being given by
\beq (z;q)_n =
\begin{cases}
1, & \text{for $n=0$}, \\ \prod_{j=0}^{n-1} (1-zq^j), & \text{for $n>0$}, \\
\prod_{j=1}^{-n} (1-zq^{-j})^{-1}, & \text{for $n<0$}.
\end{cases} \eeq
For $|q|<1$, one can write
\beq (z;q)_n \
= \ \frac{(z;q)_\infty}{(zq^n;q)_\infty}.\eeq
It is also convenient to use the notation
$$
(az^{\pm1};q)_n:= (az;q)_n(az^{-1};q)_n.
$$

The bilateral basic hypergeometric series ${}_r\psi_r$ is
defined by
\beq {}_r\psi_r \!\left[ \begin{array}{c}
         a_1,a_2,\ldots,a_r \\
         b_1,b_2,\ldots,b_r
       \end{array}
;q,z \right] = \sum_{n=-\infty}^{\infty} \frac{(a_1,a_2,
\ldots,a_r;q)_n}{(b_1,b_2,\ldots,b_r;q)_n} z^n.\eeq

\section{Mirror symmetry for $d=3$ $\mathcal{N}=2$ field theory with
$U(1)$-gauge group and $N_f=1$}\label{Nf_1}

Here we discuss consequences of the mirror symmetry for
three-dimensional $\mathcal{N}=2$ supersymmetric field theory with
$U(1)$ gauge group and $N_f=1$ in its IR fixed point, whose mirror
partner is the free Wess--Zumino theory
\cite{Intriligator:1996ex,deBoer:1997ka,Aharony:1997bx}. The initial
electric theory has one flavour or two quark superfields $Q$ and
$\tilde Q$ of charges $+1$ and $-1$ and $R$-charges
$\Delta_{Q}=\Delta_{\widetilde{Q}}=1/3$. The superconformal index is
given by the expression \beq I_{e,N_f=1} = \sum_{k \in \mathbb{Z}}
q^{|k|/3} \int_{\T}
\frac{(q^{5/6+|k|/2}z^{\pm1};q)_\infty}{(q^{1/6+|k|/2}z^{\pm1};q)_\infty}
\frac{dz}{2 \pi \textup{i} z}, \label{I1}\eeq where $\T$ is the unit
circle with positive orientation and $|q|<1$. It coincides with the
function which appeared in \cite{Imamura:2011su} as formula (67)
after the change of the chemical potential, the integration
variable, and the summation variable, given by
$$
q=x^2, \ \ \ z=e^{i a}, \ \ \ k=2s.
$$
After evaluating the integral as a sum of residues, we can write
\beq
I_{e,N_f=1} = \sum_{k \in
\mathbb{Z}} q^{|k|/3}
\frac{(q^{1+|k|},q^{2/3};q)_\infty}{(q,q^{1/3+|k|};q)_\infty}
\, {}_2
\phi_1 \!\left[ \begin{array}{c}
  q^{1/3+|k|},q^{1/3} \\
  q^{1+|k|}
\end{array} ; q,q^{2/3} \right].
\label{I1series}\eeq

The superconformal index of the mirror partner theory containing one
meson $M=Q\widetilde{Q}$ and two singlets $V_{\pm}$
\cite{Aharony:1997bx} with $R$-charge $\Delta_{M,V_{\pm}}=2/3$ has the form
 \beq I_{m,N_f=1} =
\frac{(q^{2/3};q)_\infty^3}{(q^{1/3};q)_\infty^3}.
\label{I2}\eeq
Actually, in both of these theories there exists an additional
$U(1)_J$-symmetry group \cite{Aharony:1997bx}, which is not taken into
account in \cite{Imamura:2011su}. In general, it is necessary to
introduce an additional chemical potential $w $ for it.
In the Appendix, we present these more general SCIs. The
expressions given above correspond to the choice $w =1$.

Mirror symmetry is supposed to lead to equal superconformal indices of two
theories. In \cite{Imamura:2011su}, coincidence of only the first several terms
of the series expansions in $q$ of \eqref{I1} and \eqref{I2} was checked.
Our goal is to give an analytic proof of the equality of these indices, which
we formulate as a mathematical theorem.

\begin{theorem} \label{thm:1}
The equality $I_{e,N_f=1}=I_{m,N_f=1}$ holds true.
\end{theorem}

We shall present the proof of a more general statement containing an
additional free parameter describing the $R$-charge of a chiral
superfield $\Delta_{\Phi}=h$. For this theory, the SCI
is given by the expression
\beq \label{SCI1fl} I_{e,N_f=1;h} = \sum_{k \in
\mathbb{Z}} a^{|k|/2} \int_{\T} \frac{(a^{1/2} q^{1/2+|k|/2}
z^{\pm1};q)_\infty}{(a^{-1/2}q^{1/2+|k|/2}z^{\pm1};q)_\infty}
\frac{dz}{2 \pi \textup{i} z}, \eeq
where we write $a=q^{1-h}$ and assume the constraint $|q/a|<1$.
Note that from first glance one cannot replace
in the series summand the modulus $|k|$ by $k$ since then, for any $a$,
there will be a negative value of $k$ such that the contour $\T$
stops to separate the geometric sequences of poles converging to
zero from their reciprocals.

After evaluating this integral as a sum of residues, one finds
\beq
I_{e,N_f=1;h} = \sum_{k \in \mathbb{Z}} a^{|k|/2}
\frac{(q^{1+|k|},a;q)_\infty}{(q,q^{1+|k|}/a;q)_\infty} \,
{}_2 \phi_1 \!\left[ \begin{array}{c}
  q^{1+|k|}/a,q/a \\
  q^{1+|k|}
\end{array} ; q, a \right].
\label{I3}\eeq
The superconformal index of the mirror theory (which
is again a free theory of chiral superfields) will be
\beq
I_{m,N_f=1;h} =
\frac{(a;q)_\infty}{(q/a;q)_\infty}
\frac{(q/a^{1/2};q)_\infty^2}{(a^{1/2};q)_\infty^2}.
\label{I3m}\eeq
Again these expressions correspond to the choice $w =1$ of the
chemical potential associated with the $U(1)_J$-group (see the Appendix).

\begin{theorem} \label{thm:2}
The equality $I_{e,N_f=1;h}=I_{m,N_f=1;h}$ holds true.
\end{theorem}

We start by evaluating the double sum
\begin{equation}
\sum _{k={-\infty}} ^{\infty} \frac {{{a^{{k/ 2}}}\,({\let \over /
  \def\frac#1#2{#1 / #2} a}; q) _{\infty} \,({\let \over / \def\frac#1#2{#1 /
  #2} {q^{1 + k}}}; q) _{\infty} }
} {({\let \over / \def\frac#1#2{#1 / #2} q}; q)
  _{\infty} \,({\let \over / \def\frac#1#2{#1 / #2} {{{q^{1 + k}}}\over a}};
  q) _{\infty} }
{ \sum _{n=0} ^{\infty} {\frac {({\let \over
  / \def\frac#1#2{#1 / #2} {q\over a}}; q) _{n} \,({\let \over /
  \def\frac#1#2{#1 / #2} {{{q^{1 + k}}}\over a}}; q) _{n} } {({\let \over
  / \def\frac#1#2{#1 / #2} q}; q) _{n} \,({\let \over / \def\frac#1#2{#1 / #2}
  {q^{1 + k}}}; q) _{n} }}}{a^n}.
\label{eq:1}
\end{equation}

In order to do this, we interchange the sums over $k$ and $n$ and
write the (now) inner sum over $k$ in bilateral series notation:
$$
\sum _{n=0} ^{\infty} {\frac{{a^n}\,{{({\let \over / \def\frac#1#2{#1 /
#2} {q\over
  a}}; q) _{n} ^2}}\,({\let \over / \def\frac#1#2{#1 / #2} a}; q) _{\infty}
  } {{{({\let \over / \def\frac#1#2{#1 / #2} q}; q) _{n}^2
  }}\,({\let \over / \def\frac#1#2{#1 / #2} {q\over a}}; q) _{\infty} }}
{} _{1} \psi _{1} \! \left [              \begin{matrix} \let \over
/
  \def\frac#1#2{#1 / #2} {{{q^{1 + n}}}\over a}\\ \let \over /
  \def\frac#1#2{#1 / #2} {q^{1 + n}}\end{matrix} ;q, {\displaystyle {a^{1/2}}}
  \right ].
$$
The $_1\psi_1$-series can be summed by means of Ramanujan's
summation (see \cite[(5.2.1)]{GaRaAA})
\begin{equation}
{} _{1} \psi _{1} \! \left [ \begin{matrix} \let \over / A\\ \let
\over / B\end{matrix}
   ;q, {\displaystyle Z} \right ] =
  {\frac {(\let \over / q, {B\over A}, A Z, {q\over {A Z}} ;q) _\infty}
   {(\let \over / B, {q\over A}, Z, {B\over {A Z}} ;q) _\infty} }.
\label{rama}\end{equation}
If we apply this formula and then write the sum over $n$ in basic
hypergeometric notation, then we obtain the expression
$$
{\frac{
     ({\let \over / a}; q) _{\infty} \,
     ({\let \over / {q\over {{a^{1/2}}}}}; q)
      _{\infty} }
   {({\let \over / {a^{1/2}}}; q) _{\infty} \,
     ({\let \over /{q\over a}}; q) _{\infty} }}
{} _{1} \phi _{0} \! \left [             \begin{matrix} \let \over /
      \def\frac#1#2{#1 / #2} {q\over a}\\ \let \over / \def\frac#1#2{#1 / #2}
      -\end{matrix} ;q, {\displaystyle {a^{1/2}}} \right ].
$$
After application of the $q$-binomial theorem (see \cite[(1.3.2)]{GaRaAA})
$$
{}_1\phi _0\!\left [ \begin{matrix} \let\over/ A\\ \let\over/
-\end{matrix} ;q,Z\right ] =  \frac {{( \let\over/ A Z;q)_\infty}}
{{( \let\over/  Z;q)_\infty}},
$$
our expression simplifies to \eqref{I3m}.

Now, the double sum \eqref{eq:1} was not exactly what we wanted. We
want
\begin{equation} \label{eq:3}
\sum _{k={-\infty}} ^{\infty} \frac {{{a^{\frac{\vert k\vert}
2}}\,({\let \over /
  \def\frac#1#2{#1 / #2} a}; q) _{\infty} \,({\let \over / \def\frac#1#2{#1 /
  #2} {q^{1 + \vert k\vert}}}; q) _{\infty} }
} {({\let \over / \def\frac#1#2{#1 / #2} q}; q)
  _{\infty} \,({\let \over / \def\frac#1#2{#1 / #2} {{{q^{1 + \vert k\vert}}}\over a}};
  q) _{\infty} }
{ \sum _{n=0} ^{\infty} {\frac{({\let \over
  / \def\frac#1#2{#1 / #2} {q\over a}}; q) _{n} \,({\let \over /
  \def\frac#1#2{#1 / #2} {{{q^{1 + \vert k\vert}}}\over a}}; q) _{n} } {({\let \over
  / \def\frac#1#2{#1 / #2} q}; q) _{n} \,({\let \over / \def\frac#1#2{#1 / #2}
  {q^{1 + \vert k\vert}}}; q) _{n} }}}{a^n},
\end{equation}
which is just the expression in \eqref{I3}, where the
${}_2\phi_1$-series is written out explicitly.

So, in view of the above, it suffices to prove that the sum over
non-negative $k$ in \eqref{eq:1} equals the sum over non-positive
$k$ in \eqref{eq:1}, that is,
\begin{multline} \label{eq:4}
\sum _{k={0}} ^{\infty} \frac {{{a^{{\frac k 2}}}\,({\let \over /
   a}; q) _{\infty} \,({\let \over /  {q^{1 + k}}}; q) _{\infty} }
} {({\let \over /  q}; q)
  _{\infty} \,({\let \over /  {{{q^{1 + k}}}\over a}};
  q) _{\infty} }
{ \sum _{n=0} ^{\infty} {\frac{({\let \over
  /  {q\over a}}; q) _{n} \,({\let \over /
   {{{q^{1 + k}}}\over a}}; q) _{n} } {({\let \over
  /  q}; q) _{n} \,({\let \over /
  {q^{1 + k}}}; q) _{n} }}}{a^n}\\
=\sum _{k={-\infty}} ^{0} \frac {{{a^{{\frac k 2}}}\,({\let \over /
   a}; q) _{\infty} \,({\let \over /  {q^{1 + k}}}; q) _{\infty} }
} {({\let \over /  q}; q)
  _{\infty} \,({\let \over /  {{{q^{1 + k}}}\over a}};
  q) _{\infty} }
{ \sum _{n=0} ^{\infty} {\frac{({\let \over
  /  {q\over a}}; q) _{n} \,({\let \over /
   {{{q^{1 + k}}}\over a}}; q) _{n} } {({\let \over
  /  q}; q) _{n} \,({\let \over /
  {q^{1 + k}}}; q) _{n} }}}{a^n}.
\end{multline}
This is not too difficult to see: we take the right-hand side of
\eqref{eq:4} and replace $k$ by $-k$:
$$
\sum _{k={0}} ^{\infty} \frac {{{a^{-{\frac k 2}}}\,({\let \over /
   a}; q) _{\infty} \,({\let \over /  {q^{1 - k}}}; q) _{\infty} }
} {({\let \over /  q}; q)
  _{\infty} \,({\let \over /  {{{q^{1 - k}}}\over a}};
  q) _{\infty} }
{ \sum _{n=0} ^{\infty} {\frac{({\let \over
  /  {q\over a}}; q) _{n} \,({\let \over /
   {{{q^{1 - k}}}\over a}}; q) _{n} } {({\let \over
  /  q}; q) _{n} \,({\let \over /
  {q^{1 - k}}}; q) _{n} }}}{a^n}.
$$
We see now that, actually, this is an undefined expression, due to
the terms $({\let \over /  {q^{1 - k}}}; q) _{\infty}$ and $({\let
\over /  {q^{1 - k}}}; q) _{n}$ in numerator and denominator,
respectively. We have to interpret this as an appropriate limit. In particular,
the terms in the second sum for $n=0,1,\dots,k-1$ do not contribute
anything. We may therefore start the sum over $n$ at $n=k$. This
leads to the expression
\begin{multline*}
\sum _{k={0}} ^{\infty} \frac {{{a^{-{\frac k 2}}}\,({\let \over /
   a}; q) _{\infty} \,({\let \over /  {q^{1 - k}}}; q) _{\infty} }
} {({\let \over /  q}; q)
  _{\infty} \,({\let \over /  {{{q^{1 - k}}}\over a}};
  q) _{\infty} }
{ \sum _{n=k} ^{\infty} {\frac {({\let \over
  /  {q\over a}}; q) _{n} \,({\let \over /
   {{{q^{1 - k}}}\over a}}; q) _{n} } {({\let \over
  /  q}; q) _{n} \,({\let \over /
  {q^{1 - k}}}; q) _{n} }}}{a^n}\\
=\sum _{k={0}} ^{\infty} \frac {{{a^{{\frac k 2}}}\,({\let \over /
   a}; q) _{\infty} \,({\let \over /  {q}^{1+k}}; q) _{\infty} }
} {({\let \over /  q}; q)
  _{\infty} \,({\let \over /  {{{q}^{1+k}}\over a}};
  q) _{\infty} }
{ \sum _{n=0} ^{\infty} {\frac{({\let \over
  /  {q^{1+k}\over a}}; q) _{n} \,({\let \over /
   {{{q}}\over a}}; q) _{n} } {({\let \over
  /  q^{1+k}}; q) _{n} \,({\let \over /
  {q}}; q) _{n} }}}{a^n},
\end{multline*}
where we replaced $n$ by $n+k$ and did some simplification. This is
exactly the left-hand side of \eqref{eq:4}. This proves equality of
\eqref{eq:1} and \eqref{eq:3},
and thus that the double sum \eqref{eq:3} equals the
product in \eqref{I3m}.

\textit{Remark.} The possibility to replace $|k|$ by $k$ in the
bilateral series requires some physical explanation. The variable
$k$ is the quantized magnetic flux associated with the Dirac
monopole solution in this model. It is related somehow with the spin
variable $j=|k|$ of the spherical harmonics, and the situation looks
like that after replacing $|k|$ by $k$ we come to the sum over
$j\in\Z/2$ related with non-negative values of the $SO(3)$ Casimir
operator proportional to $(j+1/2)^2$.

Note that, in contrast to $4d$ $\mathcal{N}=1$ SYM theories,
where one can often fix $U(1)_R$ group hypercharges in the infrared
fixed point, in $3d$ theories this is not the case \cite{Aharony:1997bx}.
The proven equality of SCIs remains true for arbitrary $h$ (or $a$)
although the physically acceptable theory corresponds to the value $h=1/3$
(such a possibility was  pointed out already in \cite{Imamura:2011su}).
The exact superconformal $R$-symmetry charges are found
with the help of techniques described in \cite{Jafferis:2010un}.
Thus, similar to the $4d$ situation \cite{SV1}, mathematical properties
of SCIs do not distinguish unitary theories from the non-unitary ones.
SCIs count topological objects of the theories irrespective whether
the cohomology space of $\mathcal{Q}$ and $\mathcal{Q}^\dag=\mathcal{S}$
contains ghost fields violating unitarity or not.
The equality of SCIs does not correspond necessarily to
duality/mirror symmetry between physically acceptable theories,
one should analyze separately their physical content.

\section{Mirror symmetry for $d=3$ $\mathcal{N}=2$ theory
with $U(1)$ gauge group and $N_f=2$.}\label{Nf_2}

We consider now SCIs for $3d$ $\mathcal{N}=2$ field theory with
$U(1)$ gauge group and $N_f=2$ flavours and its magnetic
partner which are dual to each other in their respective IR fixed points
\cite{Intriligator:1996ex,deBoer:1997ka,Aharony:1997bx}. The
electric theory has $F=SU(2)_l \times SU(2)_r \times U(1)_J$ flavour
symmetry for superfields with $R$-charges
$\Delta_{Q_i}=\Delta_{\widetilde{Q_i}}=h,\, i=1,2$
\cite{Aharony:1997bx}.
For each symmetry group one should introduce
chemical potentials: $u$ for $SU(2)_l$, $v$ for $SU(2)_r$ and $w $
for $U(1)_J$, but we consider first the restricted region of these
parameters where $u=v$ and $w =1$, which was analyzed perturbatively
in \cite{Imamura:2011su}. In the Appendix we present general
expressions for indices. The restricted SCI has the form \beq
\label{Nf2el} I_{e,N_f=2;h} = \sum_{k \in \mathbb{Z}} a^{|k|}
\int_{\T}
\frac{(a^{1/2}q^{1/2+|k|/2}v^{\pm1}z^{\pm1};q)_\infty}{(a^{-1/2}q^{1/2+|k|/2}
v^{\pm1}z^{\pm1};q)_\infty} \frac{dz}{2 \pi \textup{i} z},\eeq where
the chemical potential $v$ is associated with the identified $SU(2)$
flavour groups. After residue calculus, we obtain
\begin{multline}  I_{e,N_f=2;h} = \sum_{k \in \mathbb{Z}} a^{|k|}
\frac{(q^{1+|k|},q^{1+|k|}v^2,a,av^{-2};q)_\infty}{(v^{-2},
q^{1+|k|}/a, q^{1+|k|}v^2/a, q;q)_\infty} \\
\times {}_4 \phi_3 \!\left[
\begin{array}{c}
  q^{1+|k|}/a, q^{1+|k|}v^2/a, q/a, qv^2/a \\
  q^{1+|k|}, q^{1+|k|}v^2, qv^2
\end{array} ; q, a^{2} \right] + \left(v \mapsto \frac 1v\right),
\label{INf2}
\end{multline}
where $(v \mapsto \frac 1v)$ stands for the preceding
expression in which $v$ was replaced by $1/v$.
The mirror partner of this theory has again the
same $U(1)$ gauge group and two flavours with $R$-charges
$\Delta_{q_i}=\Delta_{\widetilde{q}_i}=1-h,\, i=1,2,$ and meson
fields $M_{ij}=Q_i\widetilde{Q}_j, i,j=1,2,$ with $R$-charge
$\Delta_{M_{ij}} = \Delta_{Q_i} + \Delta_{\widetilde{Q}_j} = 2h$, and
two singlet superfields $V_{\pm}$ with $R$-charge
$\Delta_{V_{\pm}}=2(1-h)$. Its restricted superconformal index has the form
 \beq I_{m,N_f=2;h}
\label{Nf2mag} =
\frac{(av^{\pm2};q)_\infty}{(qv^{\pm2}/a;q)_\infty}  \sum_{k
\in \mathbb{Z}} (q/a)^{|k|} \int_{\T}
\frac{(a^{-1/2}q^{1+|k|/2}v^{\pm1}z^{\pm1};q)_\infty}
{(a^{1/2}q^{|k|/2}v^{\pm1}z^{\pm1};q)_\infty} \frac{dz}{2 \pi
\textup{i} z}.\eeq
Note that in the generic case considered in the Appendix
the prefactor in front of the sum is considerably more complicated.
Residue calculus leads to
\begin{multline}
I_{m,N_f=2;h} =
\frac{(av^{\pm2};q)_\infty}{(qv^{\pm2}/a;q)_\infty}  \times
\left( \sum_{k \in \mathbb{Z}} (q/a)^{|k|}
\frac{(q^{1+|k|},q^{1+|k|}v^2,q/a,qv^{-2}/a;q)_\infty}{(v^{-2},
aq^{|k|}, aq^{|k|}v^2, q;q)_\infty} \right. \\
 \left. \times {}_4 \phi_3 \!\left[
\begin{array}{c}
  aq^{|k|}, aq^{|k|}v^2, a, av^2 \\
  qv^2, q^{1+|k|}, q^{1+|k|}v^2
\end{array} ; q,(q/a)^2 \right] + \left(v \mapsto \frac 1v\right) \right).
\label{INf2m}
\end{multline}

\begin{theorem} \label{thm:3}
The equality
$I_{e,N_f=2;h}=I_{m,N_f=2;h}$ holds true.
\end{theorem}

To prove this statement, we start with the double sum
\begin{multline} \label{eq:5}
\sum _{k={-\infty}} ^{\infty} \frac {a^{\vert k\vert}\,(
  q^{1+\vert k\vert},
   {q^{1 + \vert k\vert}}v^2,
   a,    a/v^2; q) _{\infty} }
 {({ 1/v^2},
 {{{q^{1 + \vert k\vert}}}/ a},
   {q^{1 + \vert k\vert}}v^2/a,
   q; q) _{\infty}}\\ \cdot
{ \sum _{n=0} ^{\infty} {\frac{
   (
  q^{1 + \vert k\vert}/ a,
  q^{1 + \vert k\vert}v^2/ a,
  q/ a,
  q v^2/ a; q) _{n}
  }
{(
  q^{1 + \vert k\vert},
  q^{1 + \vert k\vert}v^2,
  q v^2,
   q; q) _{n}
}}}{a^{2n}},
\end{multline}
representing the first term in \eqref{INf2} (the second being obtained by
reflection $v\to 1/v$). It is again easy to see that, for $k<0$, we have
\begin{multline*}
\frac {a^{-k}\,(
  q^{1-k},
   {q^{1 -k}}v^2,
   a,
   a/v^2; q) _{\infty} }
 {({ 1/v^2},
 {{{q^{1 -k}}}/ a},
   {q^{1 -k}}v^2/a,
   q; q) _{\infty}}
{ \sum _{n=0} ^{\infty} {\frac{
   (
  q^{1 -k}/ a,
  q^{1 -k}v^2/ a,
  q/ a,
  q v^2/ a; q) _{n}
  }
{(
  q^{1 -k},
  q^{1 -k}v^2,
  q v^2,
   q; q) _{n}
}}}{a^{2n}}\\
=\frac {a^{k}\,(
  q^{1+k},
   {q^{1 + k}}v^2,
   a,
   a/v^2; q) _{\infty} }
 {({ 1/v^2},
 {{{q^{1 + k}}}/ a},
   {q^{1 + k}}v^2/a,
   q; q) _{\infty}}
{ \sum _{n=0} ^{\infty} {\frac{
   (
  q^{1 + k}/ a,
  q^{1 + k}v^2/ a,
  q/ a,
  q v^2/ a; q) _{n}
  }
{(
  q^{1 + k},
  q^{1 + k}v^2,
  q v^2,
   q; q) _{n}
}}}{a^{2n}},
\end{multline*}
if one interprets the left-hand side as the appropriate limit
(namely that\linebreak $(q^ {1-k};q)_\infty/(q^ {1-k};q)_n=(q^
{1+n-k};q)_\infty$), by observing that the terms for
$n=0,1,\dots,\break k-1$ do not contribute to the sum on the left-hand side
so that one can replace $n$ by $n+k$ there. Consequently, we may
rewrite \eqref{eq:5} as
\begin{align}
\sum _{n=0} ^{\infty} \sum _{k=-n} ^{\infty} & \frac {a^{k}\,(
  q^{1+k+n},
   {q^{1 + k}}v^2,
   a,
   a/v^2; q) _{\infty} }
 {({ 1/v^2},
 {{{q^{1 + k}}}/ a},
   {q^{1 + k}}v^2/a,
   q; q) _{\infty}}
{{\frac{   (
  q^{1 + k}/ a,
  q^{1 + k}v^2/ a,
  q/ a,
  q v^2/ a; q) _{n}
  }
{(
  q^{1 + k}v^2,
  q v^2,
   q; q) _{n}
}}}{a^{2n}} \nonumber \\
&= \sum _{n=0} ^{\infty} \frac {a^{n}\,(
  q^{1+n},
   {q^{1+n} v^2,
   a,
   a/v^2; q) _{\infty} }}
 {({ 1/v^2},
 {{{q^{1+n}/ a},
   {q^{1+n} v^2/a,
   q; q) _{\infty}}
}}} {}_2\phi _1\!\left [ \begin{matrix} q v^2/a,q/a\\ q
v^2\end{matrix} ;q,
a\right ] \nonumber \\
&= \frac {(
   {q v^2,
   a,
   a/v^2; q) _{\infty} }}
 {({ 1/v^2},
 {{{q/ a},
   {q v^2/a; q) _{\infty}}
}}} {}_2\phi _1\!\left [ \begin{matrix} q v^2/a,q/a\\ q
v^2\end{matrix} ;q, a\right ] ^2. \label{eq:6}
\end{align}

Next we consider the first term in expression \eqref{INf2m}, namely
\begin{multline} \label{eq:7}
\frac {(a v^2,a/v^2;q)_\infty} {(q v^2/a,q/av^2;q)_\infty} \sum
_{k={-\infty}} ^{\infty} \frac {(q/a)^{\vert k\vert}\,(
  q^{1+\vert k\vert},
   {q^{1 + \vert k\vert}}v^2,
   q/a,
   q/av^2; q) _{\infty} }
 {({ 1/v^2},
 {{{q^{ \vert k\vert}}}a},
   {q^{ \vert k\vert}}a v^2,
   q; q) _{\infty}}\\
\times { \sum _{n=0} ^{\infty} {\frac{
   (
  q^{ \vert k\vert} a,
  q^{ \vert k\vert}a v^2,
  a,
  a v^2; q) _{n}
  }
{(
  q^{1 + \vert k\vert},
  q^{1 + \vert k\vert}v^2,
  q v^2,
   q; q) _{n}
}}}{(q/a)^{2n}}.
\end{multline}
Proceeding in the same manner as before, we see that this expression
equals
\begin{equation} \label{eq:8}
\frac {(a/v^2,
   {q v^2,
   q/a; q) _{\infty} }}
 {( q v^2/a, { 1/v^2},
 {{{a} ; q) _{\infty}
}}} {}_2\phi _1\!\left [ \begin{matrix} a v^2,a\\ q v^2\end{matrix}
;q, \frac {q} {a}\right ] ^2.
\end{equation}
The equality between \eqref{eq:6} and \eqref{eq:8} (and, hence,
between \eqref{eq:5} and \eqref{eq:7}) follows immediately from
Heine's $_2\phi_1$-transformation formula (see
\cite[(1.4.1)]{GaRaAA})
\begin{equation} \label{eq:Heine1}
{}_2\phi _1\!\left [ \begin{matrix} A,B \\ C \end{matrix}
;q, Z\right ]=\frac{(B,A Z;q)_\infty}{(C, Z;q)_\infty}
{}_2\phi _1\!\left [ \begin{matrix} C/B,Z \\ AZ\end{matrix}
;q, B\right ]
\end{equation}
after substituting $A=qv^2/a, B=q/a, C=qv^2, Z=a$.
Since the second terms in \eqref{INf2} and \eqref{INf2m} arise from
the respective first terms by the substitution $v\mapsto 1/v$,
this completes the proof of the claimed equality between SCIs.

\medskip

The full symmetry group of the ${}_2\phi_1$-series is generated by
repeated applications of this transformation together with the
permutation of its numerator parameters $A$ and $B$, and it is
isomorphic to the dihedral group $D_6$
\cite{Krattenthaler}. This leads to two further
$_2\phi_1$-transformation formulas, namely (see \cite[(1.4.5)]{GaRaAA})
\begin{equation} \label{eq:Heine2}
{}_2\phi _1\!\left [ \begin{matrix} \let\over/ A,B\\ \let\over/  C\end{matrix} ;q,Z\right ]
=   \frac {{(C/B,BZ;q)_\infty}}
{{( C,Z;q)_\infty}}
   {}_2\phi _1\!\left [ \begin{matrix} ABZ/C,B\\ BZ\end{matrix} ;q,
{\frac C B}\right ]
\end{equation}
and (see \cite[(1.4.6)]{GaRaAA})
\begin{equation} \label{eq:Heine3}
{}_2\phi _1\!\left [ \begin{matrix} \let\over/ A,B\\ \let\over/  C\end{matrix} ;q,Z\right ]
=   \frac {{( \let\over/ {{A B Z}\over C};q)_\infty}}
{{( \let\over/  Z;q)_\infty}}
   {}_2\phi _1\!\left [ \begin{matrix} \let\over/ {C\over A},{C\over B}\\ \let\over/  C\end{matrix} ;q,{\frac{A B
    Z} C}\right ].
\end{equation}
Therefore we have two more different representations of the SCI:
\begin{align} \notag
 I_{e,N_f=2;h}&=\frac {(qv^2/a,a/v^2,q,q; q) _{\infty} }
 {(1/v^2,qv^2,a,q/a; q) _{\infty}}
{}_2\phi _1\!\left [ \begin{matrix} a,a\\ q \end{matrix} ;q, \frac
{qv^2} {a}\right ] ^2 + \left(v \mapsto \frac 1v\right)
\\
&=\frac {(a/v^2; q)_{\infty}(av^2,q;q)_\infty^2 }
 {(1/v^2,q v^2, a,q/a,qv^2/a ;q) _{\infty}
} {}_2\phi _1\!\left [ \begin{matrix} q/a,q/a\\ q\end{matrix} ;q,
av^2\right ] ^2 + \left(v \mapsto \frac 1v\right),
\label{eq:8add}
\end{align}
following from the first transformation with
$A=q/a, B=qv^2/a$ and the second one with
$A=qv^2/a, B=q/a$, where we choose $C=qv^2, Z=a$ in both cases.
The third transformation does not yield new results.
It is necessary to clarify whether these two new cases can attain a
proper interpretation as SCIs of some new mirror field theories.

Expressions (\ref{eq:6}) and (\ref{eq:8}) are given by the product
of two terms coming from the sum over the monopoles and the
integration over the gauge group. This resembles a product of the
holomorphic and anti-holomorphic parts of the four-point function
of $2d$ conformal field theory, and the mirror maps resemble the
transformation formulas for these functions in different channels.

One can consider the equality $I_{e,N_f=2;h}=I_{m,N_f=2;h}$ in the limit
$v=\textup{i}$, which can be interpreted as the reduction of the
number of flavours from $N_f=2$ to $N_f=1$. For $v=\textup{i}$, the
expression $I_{e,N_f=2;h}$ reduces to $I_{e,N_f=1;h}$, where $q$ is replaced
by $q^2$ and the integration variable $z$ is replaced by $-z^2$.
The integral $I_{m,N_f=2;h}$
reduces also to $I_{e,N_f=1;h}$ with $q/a$ replaced by $a$ and with
the additional multiplier $(-a;q)^2_\infty/(-q/a;q)^2_\infty$. Both
integrals can be evaluated exactly, as described in the previous
section, and we find that the equality
$I_{e,N_f=2;h}|_{v=i}=I_{m,N_f=2;h}|_{v=i}$ is reduced to the
identity \beq
\frac{(a^{2},q^{2}/a,q^{2}/a^2;q^2)_\infty}{(q^2/a^2,a,a;q^2)_\infty}
= \frac{(q^2/a^2,aq,aq;q^2)_\infty}{(a^{2},q/a,q/a;q^2)_\infty}
\frac{(-a;q)_\infty^2}{(-q/a;q)_\infty^2}.\eeq
Multiplying both
sides by $(a^{2};q^2)_\infty/(q^2/a^2;q^2)_\infty$, one can see that
it reduces to the square of the relation
$$
\frac{(a^{2},q^{2}/a;q^2)_\infty}{(q^2/a^2,a;q^2)_\infty} =
\frac{(aq;q^2)_\infty}{(q/a;q^2)_\infty}
\frac{(-a;q)_\infty}{(-q/a;q)_\infty},
$$
which is easy to verify.

From the physical point of view, this is very similar to the
reduction $t_jt_k=pq$ of the elliptic hypergeometric integral
describing the superconformal index of a $4d$ $\mathcal{N}=1$ SYM
theory with $SU(2)$ gauge group and $8$ chiral superfields
\cite{SV}. The latter theory has many dual partners, and the $N_f=4$
to $N_f=3$ reduction leads to four different theories --- the
original interacting electric theory with reduced number of
flavours, two interacting magnetic theories with reduced number of
chiral superfields, some additional mesons, and different flavour
symmetries. The fourth theory was in the confined phase with free
mesons, and its index was given by an explicit infinite product. All
four forms correspond to the $W(E_6)$ Weyl group symmetry of the
elliptic beta integral evaluation formula \cite{S1}. Our $3d$ mirror
symmetric theories look similar to the first pair of the described
$4d$ dual theories, and we expect existence of other mirror
partners, whose SCIs would reduce for $v=\textup{i}$ directly to
expression \eqref{I3m}. The final comment concerns the
parameter $h$ related to the arbitrariness of the $R$-charges discussed
in the Introduction. As we see, equality of
SCIs holds true for arbitrary $h$, i.e., one does not need
the exact physical value of the $R$-charge
of the quarks in the IR fixed point.

\section{Generalization to arbitrary $N_f$}

Now we consider the electric $3d$ $\mathcal{N}=2$
supersymmetric theory with $U(1)$ gauge group and arbitrary number of
flavours $N_f$ with $R$-charges $\Delta_{Q_i}=\Delta_{\widetilde{Q}_i}=h,
\, i=1,\ldots,N_f$. The global symmetry group is $SU(N_f)_l \times
SU(N_f)_r \times U(1)_J$, so we should introduce chemical potentials
for each subgroup $s_i,\, t_i, i=1,\ldots, N_f$, and $w $. However,
we restrict to $s_i=t_i,\, i=1,\ldots, N_f, \, w =1$.
First we present the single-particle state index of this theory,
\beq \ind(a,q,t_i) =\frac{q^{1/2+|s|/2}(z+z^{-1})}{1-q}
 \sum_{i=1}^{N_f} \left( a^{-1/2}t_i -a^{1/2}t_i^{-1}\right),\eeq
from which one can easily compute the full SCI: \beq \label{Nfel}
I_{e,N_f;h} = \sum_{k \in \mathbb{Z}} a^{N_f|k|/2} \int_{\T}
\prod_{i=1}^{N_f} \frac{(a^{1/2}q^{1/2+|k|/2} t_i^{-1}
z^{\pm1};q)_\infty}{(a^{-1/2}q^{1/2+|k|/2} t_i z^{\pm1};q)_\infty}
\frac{dz}{2 \pi \textup{i} z},\eeq where $\prod_{i=1}^{N_f} t_i =1$.
On the mirror symmetric side we shall have an $(N_f-1)$-fold
multiple sum over the $N_f-1$ analogues of the variable $s$ and an
$(N_f-1)$-dimensional integral over the $N_f-1$ analogues of the
variable $z$. After the substitution $t_k=e^{2 \pi \textup{i}
k/N_f}$ in the electric theory SCI, we come to the initial $N_f=1$
SCI with different choice of the parameters. Therefore we expect
that the $(N_f-1)$-multiple sums/integrals of the mirror side also
reduce appropriately.

The $N_f\to N_f-1$ reduction in SCIs is realized similar to the $4d$
indices case \cite{SV1}. It can be summarized as follows. First
one should substitute $t_{N_f}=a^{1/2}$ so that, as is easily seen from
(\ref{Nfel}), the contribution from the $N_f$-th flavour drops out
(physically
this means that we have integrated this field out by giving
large mass to it). After
having done this, one should renormalize the parameters
$t_i$, $i=1,\ldots,N_f-1$, and $a$ as follows:
$$t_i \rightarrow a^{-1/2(N_f-1)} t_i, \ i=1,\ldots,N_f-1, \ \ \
a \rightarrow a^{(N_f-1)/N_f}.$$
This brings the SCI back to expression (\ref{Nfel}) with the
replacement of $N_f$ by $N_f-1$.

After residue calculus, we arrive at the following sum of
well-poised ${}_{2N_f} \phi_{2N_f-1}$-series:
\begin{multline} I_{e,N_f;h} = \sum_{k \in
\mathbb{Z}} a^{N_f|k|/2} \prod_{i=2}^{N_f}
\frac{(q^{1+|k|}t_1/t_i,a/t_it_1;q)_\infty}{(t_i/t_1,
q^{1+|k|}/at_1t_i;q)_\infty}
\frac{(q^{1+|k|},a/t_1^2;q)_\infty}{(q^{1+|k|}t_1^2/a, q;q)_\infty}
\\  \times {}_{2N_f} \phi_{2N_f-1}
\!\left[
\begin{array}{c}
  q^{1+|k|}\frac{t_1^2}{a}, q^{1+|k|}\frac{t_1t_2}{a}, \ldots , q^{1+|k|}
\frac{t_1t_{N_f}}{a},
\frac{qt_1^2}{a}, \frac{qt_1t_2}{a}, \ldots, \frac{qt_1t_{N_f}}{a} \\
 \frac{qt_1}{t_2}, \ldots, \frac{qt_1}{t_{N_f}}, q^{1+|k|},
q^{1+|k|} \frac{t_1}{t_2}, \ldots, q^{1+|k|} \frac{t_1}{t_{N_f}}
\end{array} ; q, a^{N_f} \right]  \\
 + \text{idem} \left[ 1; 2,\ldots,N_f
\right],
\end{multline}
where $\text{idem} \left[ 1; 2,\ldots,N_f \right]$
means that one has to add the sum of all expressions arising from
the previous expression by interchanging the
index $1$ with the indices $2,\dots,N_f$, one by one.

As in the previous section, we can rewrite the above expression in
the form
\begin{multline} \label{separ}
I_{e,N_f;h} =
\frac{(a/t_1^2,qt_1^2/a;q)_\infty}{(q,q;q)_\infty} \prod_{i=2}^{N_f}
\frac{(a/t_1t_i,qt_1t_i/a;q)_\infty}{(t_i/t_1,qt_1/t_i;q)_\infty}
\sum_{k=0}^\infty a^{N_fk}
\frac{(q^{1+k};q)^2_\infty}{(q^{1+k}t_1^2/a;q)^2_\infty}
\\
 \times \prod_{i=2}^{N_f}
\frac{(q^{1+k}t_1/t_i;q)^2_\infty}{(q^{1+k}t_1t_i/a;q)^2_\infty}
\,{}_{N_f} \psi_{N_f} \!\left[
\begin{array}{c}
  q^{1+k}t_1^2/a, q^{1+k}t_1t_2/a, \ldots , q^{1+k} t_1t_{N_f}/a \\
  q^{1+k}, q^{1+k} t_1/t_2, \ldots, q^{1+k} t_1/t_{N_f}
\end{array} ; q, a^{N_f/2} \right] \\
+ \text{idem} \left[ 1; 2,\ldots,N_f
\right],
\end{multline}
which finally is rewritten as
\begin{multline} \label{prod}
I_{e,N_f;h} =
\frac{(a/t_1^2;q)_\infty}{(qt_1^2/a;q)_\infty} \prod_{i=2}^{N_f}
\frac{(a/t_1t_i,qt_1/t_i;q)_\infty}{(t_i/t_1,qt_1t_i/a;q)_\infty}
\\
\times {}_{N_f} \phi_{N_f-1}
\!\left[
\begin{array}{c}
  q t_1^2/a, q t_1t_2/a, \ldots , q t_1t_{N_f}/a \\
  q t_1/t_2, \ldots, q t_1/t_{N_f}
\end{array} ; q, a^{N_f/2} \right] ^2
\\
+ \text{idem} \left[ 1; 2,\ldots,N_f
\right].
\end{multline}
Here, we see again the separation of the sum and the integral into
$N_f$ terms, each being the square of an
${}_{N_f}\phi_{N_f-1}$-series.

One can check that the term $|k|$
in the summand can be replaced by $k\in\mathbb{Z}$ (both before and
after residue calculus), which automatically leads to
the termination of the $k$-series from below.
Indeed, when $k < 0$ we have
\begin{align}\notag
a^{N_fk/2} \prod_{i=1}^{N_f}&
\frac{(a^{1/2}q^{1/2+k/2} t_i^{-1}
z^{\pm1};q)_\infty}{(a^{-1/2}q^{1/2+k/2} t_i z^{\pm1};q)_\infty}
\\ \notag
&= a^{N_fk/2} \prod_{i=1}^{N_f}
\frac{(a^{1/2}q^{1/2+k/2} t_i^{-1} z^{\pm1};q)_{-k}
(a^{1/2}q^{1/2-k/2} t_i^{-1}
z^{\pm1};q)_\infty}{(a^{-1/2}q^{1/2+k/2} t_i z^{\pm1};q)_{-k}
(a^{-1/2}q^{1/2-k/2} t_i z^{\pm1};q)_\infty}
\\
&= a^{-N_fk/2} \prod_{i=1}^{N_f}
\frac{(a^{1/2}q^{1/2-k/2} t_i^{-1}
z^{\pm1};q)_\infty}{(a^{-1/2}q^{1/2-k/2} t_i z^{\pm1};q)_\infty},
\end{align}
after taking into account the constraint $\prod_{i=1}^{N_f} t_i = 1$.
Thus we come to the following form of the SCI:
\beq
I_{e,N_f;h} = \sum_{k \in \mathbb{Z}} a^{N_fk/2} \int_{\T}
\prod_{i=1}^{N_f} \frac{(a^{1/2}q^{1/2+k/2} t_i^{-1}
z^{\pm1};q)_\infty}{(a^{-1/2}q^{1/2+k/2} t_i z^{\pm1};q)_\infty}
\frac{dz}{2 \pi \textup{i} z},
\label{index5}\eeq
with the same contour of integration $\mathbb{T}$ as before.

Now we would like to discuss another representation of SCIs.
Interchanging the sum over $k$ and integration over $z$ in
\eqref{index5}, we can write
\begin{multline*}
I_{e,N_f;h} = \int_{\T}
\prod_{i=1}^{N_f} \frac{(\sqrt{aq} t_i^{-1}
z^{\pm1};q)_\infty}{(\sqrt{q/a} t_i z^{\pm1};q)_\infty}\,
{}_{2N_f}\psi_{2N_f} \!\left[
\begin{array}{c}
  \sqrt{q/a} t_i z^{\pm1}
\\
  \sqrt{aq} t_i^{-1} z^{\pm1}
\end{array} ; q,a^{N_f} \right]
\frac{dz}{2 \pi \textup{i} z} \\
+
a^{N_f/2} \int_{\T} \prod_{i=1}^{N_f} \frac{(q\sqrt{a} t_i^{-1}
z^{\pm1};q)_\infty}{(qt_i z^{\pm1}/\sqrt{a};q)_\infty}\,
{}_{2N_f}\psi_{2N_f} \!\left[
\begin{array}{c}
  q t_i z^{\pm1}/\sqrt{a} \\
  q\sqrt{a} t_i^{-1} z^{\pm1}
\end{array} ; q,a^{N_f} \right]
\frac{dz}{2 \pi \textup{i} z}.
\end{multline*}

It would interesting to find a pure integral representation of
this expression, which may help in understanding a connection with
elliptic hypergeometric integrals. We were able to represent the SCI as
a sum of two integrals only in the case $N_f=1$. For that purpose, we use the
following compact contour integral representation for a general
${}_2\psi_2$-series: \beq \label{2psi2} {}_2\psi_2\!\left[
\begin{matrix} a, q/d
\\ b, q/c\end{matrix}; q, \frac{\alpha d}{\beta c}\right]
=\frac{(q,q,b/a,d/c;q)_\infty}{(b,d,q/a,q/c;q)_\infty} \int_\T
\frac{(az\alpha,cz\beta,q/az\alpha,q/cz\beta;q)_\infty}
{(z\alpha,z\beta,b/az\alpha,d/cz\beta;q)_\infty}\frac{dz}{2\pi iz},
\eeq where $|b/a|<|\alpha/\beta|<|c/d|$. It is obtained after
multiplying two $_1\psi_1$-series \eqref{rama} with different
choices of parameters $(A,B)$, say $(a,b)$ respectively $(c,d)$,
and $Z_1=\alpha
z$, $Z_2=\beta z$, with the subsequent integration $\int_\T dz/z$.
In this way we find the expression
\begin{multline} \frac{I_{e,N_f=1;h}}{
(a,q;q)^2_\infty } =\int_{\mathbb{T}^2}
\frac{(zw\sqrt{q/a},\sqrt{aq}/zw;q)^2_\infty}
{(\sqrt{aq}z^{\pm1},\sqrt{q/a}z^{\pm1};q)_\infty (w,
a/w;q)^2_\infty}
\frac{dz}{2 \pi \textup{i} z} \frac{dw}{2 \pi \textup{i} w} \\
+  \sqrt{a}  \int_{\mathbb{T}^2}
\frac{(q (zw /\sqrt{a})^{\pm1},(\sqrt{a}/zw)^{\pm1};q)_\infty}
{(\sqrt{a}z^{\pm1},q z^{\pm1}/\sqrt{a};q)_\infty(w, a/w;q)^2_\infty}
\frac{dz}{2 \pi \textup{i} z} \frac{dw}{2 \pi \textup{i} w}.
\end{multline}
The usefulness of such a representation is not clear at present.

\section{Concluding remarks}\label{concl}

We described an interplay between
mirror-symmetric three-dimensional superconformal
field theories and the theory of $q$-special functions.
This is similar to the relation between SCIs in four
dimensions and the theory of elliptic hypergeometric integrals
\cite{Dolan:2008qi,SV1}. Actually, three-dimensional
$\mathcal{N}=2$ supersymmetric field theories can be obtained by
dimensional reduction from four-dimensional $\mathcal{N}=1$
theories and thus the SCIs of these theories should be
related by some reduction procedure. So far this question,
which was the main motivation for our work, is not
understood and requires a clarification.
The calculation of $3d$ SCIs is not so straightforward as in
$4d$ theories, and the sums over $3d$ monopole fluxes
require proper interpretation in terms of the contour integrals.
An understanding of this connection would allow us to apply the full
power of $4d$ dualities together with the theory
of elliptic hypergeometric integrals defining SCIs \cite{S1}.
Another related problem is the problem of extending our results to gauge
groups other than $U(1)$ which we have considered here.

$q$-Special functions have a much poorer structure
than elliptic hypergeometric functions. For example,
for $3d$ SCIs the kernels of integral-sums satisfy first order
$q$-difference equations with rational coefficients, which is the
only substitute for the $4d$ notion of total ellipticity, which is
interpreted from the physical point of view as the
 't Hooft anomaly matching conditions \cite{SV1}. The only somewhat
analogous notion is the parity anomaly \cite{Aharony:1997bx} (see
also \cite{parity} for early references and a perturbative analysis
of this phenomenon), which is associated with the classical
Chern--Simons term breaking the parity.

As to reduction of the number of flavours, in the $4d$ case there
was a very simple procedure for doing this at the level of SCIs ---
it was sufficient to restrict parameters to $s_kt_k=pq$ for
removing the $k$-th flavour \cite{SV1}. In the $3d$ case, the
analogous $N_f=2 \; \to \; N_f=1$ result is reached in
$\mathcal{N}=2$ supersymmetric theory with $U(1)$ gauge group by the
parameter restriction $v=\textup{i}$, and a similar reduction from
arbitrary $N_f$ to $N_f=1$ should be valid after using higher roots
of unity. The reduction $N_f\to N_f-1$ in
the general case is realized by the special choice of one parameter,
namely $t_{N_f}=a^{1/2}$, with a subsequent renormalization of the
remaining $t_j$'s and $a$.

Another interesting fact we observe is the factorization of the sum
over the monopoles and the integration over the gauge group in $3d$
SCIs of $\mathcal{N}=2$ supersymmetric field theory with $U(1)$ gauge
group and $N_f$ flavours; see (\ref{prod}). It resembles much the
separation into holomorphic and antiholomorphic parts of the
correlation functions in $2d$ CFT.

\bigskip

{\bf Acknowledgments.} We are grateful to I. Bandos and F. A. H.
Dolan for valuable discussions. G.V. would like to thank H. Nicolai
for general support. V.S. thanks the Albert-Einstein-Institut in
Golm, where this paper was initiated during his visit
in February 2011, for its warm hospitality.
Additionally, V.S. and G.V. thank the Heisenberg--Landau program for
financial support.

{\bf Note added.}
After finishing this work, we became aware of \cite{N8},
where SCIs induced by the topological charge of the group $U(1)_J$
are considered for $3d$ $\mathcal{N}=8$ theories,
which partially overlaps with our considerations.
Furthermore, the revised version of \cite{Imamura:2011uj},
which appeared after publishing the preprint version of our
paper, includes perturbative consideration of SCIs related to the
group $U(1)_J$, and its authors affirm an overlap with our results.

\appendix
\section{General  SCIs}\label{chSUSP}

\subsection{$N_f=1$ case}

Let us present the formulas for $3d$ supersymmetric $U(1)$ gauge
group with one flavour for general chemical potentials: \beq
I_{e,N_f=1;h;w} = \sum_{k \in \mathbb{Z}} a^{|k|/2} w^k
\int_{\T} \frac{(a^{1/2} q^{1/2+|k|/2}
z^{\pm1};q)_\infty}{(a^{-1/2}q^{1/2+|k|/2}z^{\pm1};q)_\infty}
\frac{dz}{2 \pi \textup{i} z}. \eeq
Here, a crucial difference from
the $w=1$ index considered in the main text is that the sum
over $s$ contains both $|k|$ and $k$, which, actually, requires a
modification of the general formula found in \cite{Kim:2009wb,Imamura:2011uj}.
Namely, we suggest that this term appears after the substitution of the
term $w^Q$ into the general trace \eqref{trace}, where $Q$ is the
conserved charge of the global symmetry group $U(1)_J$ generated by the
topological current $J^\mu=\epsilon^{\mu\nu\sigma}F_{\nu\sigma}$
\cite{Aharony:1997bx} which is related to the monopole flux.

After residue calculus, we get
\beq
I_{e,N_f=1;h;w} = \sum_{k \in \mathbb{Z}} a^{|k|/2} w^k
\frac{(q^{1+|k|},a;q)_\infty}{(q,q^{1+|k|}/a;q)_\infty} \,
{}_2 \phi_1 \!\left[ \begin{array}{c}
  q^{1+|k|}/a,q/a \\
  q^{1+|k|}
\end{array} ; q, a \right].
\label{I3gen}\eeq
The superconformal index of the mirror theory (which
is again a free theory of chiral superfields) is
\beq
I_{m,N_f=1;h;w} =
\frac{(a;q)_\infty}{(q/a;q)_\infty}
\frac{(q w^{\pm1}/a^{1/2};q)_\infty}{(a^{1/2} w^{\pm1};q)_\infty}.
\label{I3mgen}\eeq
The equality $I_{e,N_f=1;h;w} = I_{m,N_f=1;h;w}$ of SCIs
is proven completely analogously to the case $w=1$. We leave the
details to the reader.

\subsection{$N_f=2$ case}

Now we consider the most general SCI for dual theories with
$N_f=2$. The initial theory index has the form \beq
I_{e,N_f=2;h;w} = \sum_{k \in \mathbb{Z}} a^{|k|} w^k
\int_{\T} \frac{(a^{1/2}q^{1/2+|k|/2}u^{\pm1}z^{-1},
a^{1/2}q^{1/2+|k|/2}v^{\pm1}z;q)_\infty}{(a^{-1/2}q^{1/2+|k|/2}
u^{\pm1}z, a^{-1/2}q^{1/2+|k|/2} v^{\pm1}z^{-1};q)_\infty}
\frac{dz}{2 \pi \textup{i} z},\eeq where the chemical potentials $u$
and $v$ are associated with the $SU(2)_l\times SU(2)_r$ flavour group
and $w$ corresponds to the $U(1)_J$-group discussed in
\cite{Aharony:1997bx}. After residue calculus, we obtain
\begin{multline}
I_{e,N_f=2;h;w} = \sum_{k\in \mathbb{Z}} a^{|k|} w^k
\frac{(q^{1+|k|},q^{1+|k|}v^2,au/v,a/uv;q)_\infty}{(v^{-2},
q^{1+|k|}v/au, q^{1+|k|}uv/a, q;q)_\infty}  \\
\times {}_4 \phi_3 \!\left[
\begin{array}{c}
  q^{1+|k|}v/au, q^{1+|k|}uv/a, qv/au, quv/a \\
  q^{1+|k|}, q^{1+|k|}v^2, qv^2
\end{array} ; q, a^{2} \right] + \left(v \mapsto \frac 1v\right),
\end{multline}
which gives \beq I_{e,N_f=2;h;w} = \frac {(
   q v^2,
   au/v,
   a/uv; q) _{\infty} }
 {({ 1/v^2},
 {{{qv/aa},
   {q uv/a; q) _{\infty}}
}}} {}_2\phi _1\!\left [ \begin{matrix} q uv/a,qv/au\\ q
v^2\end{matrix} ;q, a w^{\pm1} \right ]  + \left(v \mapsto
\frac 1v\right).\label{general_chem}\eeq
Here, we use again a short notation: the term $w^{\pm1}$ indicates
that we actually mean the product of two $_2\phi_1$-series, one
with $w$ in place of $w^{\pm1}$, the other with $w^{-1}$ in place of
$w^{\pm1}$.

The mirror partner has SCI of the form
\begin{multline}
I_{m,N_f=2;h;w} = \frac{(q w^{\pm1}/a;q)_\infty}{(a w^{\pm1};q)_\infty}
 \frac{(a u^{\pm1} v^{\pm1};q)_\infty}{(q u^{\pm1} v^{\pm1}/a;q)_\infty} \\
\times \sum_{k \in \mathbb{Z}} (q/a)^{|k|} w^k \int_{\T}
\frac{(a^{-1/2}q^{1+|k|/2}v^{\pm1}z^{-1},
a^{-1/2}q^{1+|k|/2}u^{\pm1}z;q)_\infty} {(a^{1/2}q^{|k|/2}v^{\pm1}z,
a^{1/2}q^{|k|/2}u^{\pm1}z^{-1};q)_\infty} \frac{dz}{2 \pi \textup{i}
z}.\end{multline}
The position of some of the poles of the integrand are
proportional to $u$. To consider the poles with position
proportional to $v$ one can make the change $z \mapsto 1/z$, and then
we have a situation similar to the one discussed in the main part
of the text. Residue calculus leads to
\begin{multline}
I_{m,N_f=2;h;w} = \frac{(q w^{\pm1}/a;q)_\infty}{(a w^{\pm1};q)_\infty}
\frac{(a u^{\pm1} v^{\pm1};q)_\infty}{(q u^{\pm1} v^{\pm1}/a;q)_\infty} \\
\times \left( \sum_{k \in \mathbb{Z}} (q/a)^{|k|} w^k
\frac{(q^{1+|k|},q^{1+|k|}v^2,qu/av,q/auv;q)_\infty}{(v^{-2},
aq^{|k|}v/u, aq^{|k|}uv, q;q)_\infty} \right. \kern3cm\\
 \left. \makebox[-1em]{} \times {}_4 \phi_3 \!\left[
\begin{array}{c}
  aq^{|k|}v/u, aq^{|k|}uv, av/u, auv \\
  q^{1+|k|}, q^{1+|k|}v^2, qv^2
\end{array} ; q,(q/a)^2 \right] + \left(v \mapsto \frac 1v\right)
\right),
\end{multline}
and, finally, to
\begin{multline}
I_{m,N_f=2;h;w} = \frac{(q w^{\pm1}/a;q)_\infty}{(a w^{\pm1};q)_\infty}
\frac{(a u^{\pm1} v^{\pm1};q)_\infty}{(qu^{\pm1} v^{\pm1}/a;q)_\infty} \\
\times \left( \frac {(
   {q v^2,
   qu/av,
   q/auv; q) _{\infty} }}
 {({ 1/v^2},
 {{{av/u},
   {auv; q) _{\infty}}
}}} {}_2\phi _1\!\left [ \begin{matrix} auv,av/u\\ q v^2\end{matrix}
;q, q w^{\pm1}/a \right ]  + \left(v \mapsto \frac
1v\right) \right),\label{general_chem2}
\end{multline}
with the same short notation as in \eqref{general_chem}.

One can easily see that (\ref{general_chem}) and
(\ref{general_chem2}) coincide by using the transformation formulas
\eqref{eq:Heine1} and \eqref{eq:Heine3} in the text. Thus we established
the equality of SCIs in the most general possible setting.

\end{document}